\documentclass[paper]{ieice}
\usepackage{booktabs}
\usepackage{graphicx}
\usepackage{nicefrac}
\usepackage[T1]{fontenc}
\usepackage{soul}

\usepackage{xspace}
\usepackage{flushend}  
\usepackage{color}
\usepackage{url}
\usepackage{booktabs}
\usepackage[table,xcdraw]{xcolor}
\usepackage{multirow}
\usepackage{listliketab}
\usepackage{amssymb}
\usepackage{algorithmic}
\usepackage{booktabs}
\usepackage{comment}
\usepackage{amsmath}
\usepackage{pgfplots}
\usepackage{bbding}
\usepackage{listings}
\usepackage{float}
\usepackage{tabularx}
\usepackage{cite}
\usepackage{rotating}
\usepackage[table,xcdraw]{xcolor}
\usepackage[square,sort,comma,numbers]{natbib}
\usepackage{tcolorbox}
\usepackage{ulem}
\usepackage{censor}
\usepackage{tikz}
\usepackage{amssymb} 
\usepackage{amsfonts}
\usepackage{graphicx,subfigure}
\usepackage{xparse}
\usepackage[colorinlistoftodos]{todonotes}
\usepackage{ulem}
\usepackage{amssymb}
\usepackage{lmodern}
\usepackage{mathtools, nccmath}
\usepackage{venndiagram}
\usepackage[symbol]{footmisc}

\usepackage{balance}
\usepackage{esvect}
\usepackage{xurl}
            
\usepackage[linesnumbered,ruled,vlined]{algorithm2e}

\SetCommentSty{mycommfont}

\SetKwInput{KwInput}{Input}                % Set the Input
\SetKwInput{KwOutput}{Output}              % set the Output

\makeatletter
\def\@xfootnote[#1]{%
  \protected@xdef\@thefnmark{#1}%
  \@footnotemark\@footnotetext}
\makeatother

\setcitestyle{square}

\newcommand{\RQOne}{\textbf{RQ$_1$: What proportion of package usage information mined from Stack Overflow exist in npm packages?}}
\newcommand{\RqTwo}{\textbf{RQ$_2$: What kinds of answers are posted in response to questions that include package usage information?}}

\newcommand\syful[1]{{\textcolor{black}{#1}}}

\usepackage[colorinlistoftodos]{todonotes}

\field{D}
\vol{}
\no{}
\SpecialSection{Empirical Software Engineering}
\title{An Exploration of \texttt{npm} Package Co-Usage Examples from Stack Overflow: A Case Study}

\authorlist{%
\authorentry{Syful Islam}{n}{naist}\MembershipNumber{}
\authorentry{Dong Wang}{n}{naist}\MembershipNumber{}
\authorentry{Raula Gaikovina Kula}{n}{naist}\MembershipNumber{}
\authorentry{Takashi Ishio}{m}{naist}\MembershipNumber{1301152}
\authorentry{Kenichi MATSUMOTO}{f}{naist}\MembershipNumber{8925358}
}
\affiliate[naist]{The authors are with the Graduate School of Science and Technology, Nara Institute of Science and Technology, Ikoma-shi, 630-0192, Japan.
}
\received{2021}{02}{26}
\revised{2021}{7}{14}

\begin{document}
\maketitle
\begin{summary}
\noindent Third-party package usage has become a common practice in contemporary software development. Developers often face different challenges, including choosing the right libraries, installing errors, discrepancies, setting up the environment, and building failures during software development.
The risks of maintaining a third-party package are well known, but it is unclear how information from Stack Overflow (SO) can be useful. 
This paper performed an empirical study to explore npm package co-usage examples from SO.
From over 30,000 SO question posts, we extracted 2,100 posts with package usage information and matched them against the 217,934 npm library package. 
We find that, popular and highly used libraries are not discussed as often in SO.
However, we can see that the accepted answers may prove useful, as we believe that the usage examples and executable commands could be reused for tool support. 

\end{summary}

\section{Introduction} 
Usage of third-party packages in contemporary software development has become a common practice by developers. 
For example, \texttt{npm} (i.e., Node.js package manager) is by far the largest package manager, allowing developers to reuse functionality instead of creating their own, saving both time with minimal efforts.
The \texttt{npm} ecosystem has provided over 800,000 free and reusable software libraries and is trusted by over 11 million developers around the world.\footnote[1]{\url{https://www.npmjs.com/}}

Despite these benefits of using packages, developers constantly face a variety of issues when using them.
Dietrich et al.~\cite{dietrich2014broken} performed a case study and showed that partial package upgrades have high potential to introduce binary incompatibility problems at build time.
Raemaekers et al.~\cite{raemaekers2017semantic, raemaekers2014semantic} pointed out that developers are wary of the inherent costs and risks of package incompatibilities when integrating new and unknown packages into their systems.
Most prior work have explored package usage~\cite{de2018library, dietrich2019dependency}, developing package recommendation tools~\cite{thung2013automated, ouni2017search, saied2018improving, alrubaye2020learning, yu2017combining, nguyen2020crossrec}.

Previous studies reported that question-answering websites such as Stack Overflow (SO) are useful for communicating developers' issues.
Several studies have been conducted on SO resources including source code utilization~\cite{wu2019developers}, analogical libraries recommendation~\cite{chen2016similartech}, fixing run-time exception~\cite{mahajan2020recommending}, improving API documentation~\cite{treude2016augmenting}, API usage scenarios~\cite{uddin2020mining} and so forth.
Other studies have focused on more interviews and surveys of developers~\cite{10.1145/3368089.3409711, xavier2017we}. We hypothesize that the library usage information from SO may also be beneficial for developers. While the risks of maintaining third-party libraries are well known,  it is still unclear that whether the library usage information mined from question-answering sites are useful or not in maintaining libraries.

To fill this gap, in this paper, we perform an exploratory study on package usage information from SO in term of co-usage relationship.
As defined by Todorov et. al \citep{todorov2017sol}, co-usage is the pattern of package dependencies that are used together. The rationale behind refining the co-usage relationship is to study problems caused by npm packages.  
In particular we investigate (i) whether we can detect package usage (i.e., co-usage) information from SO and (ii) what the developers are looking for to solve problems related to the package.
To address these, we study over 2,100 SO posts and matched them to 217,934 npm library packages. 
We reveal the following valuable lessons along the way: 

\textbf{Lesson 1:}
We find that only three out of the top ten of the most used npm libraries are mentioned in SO. The top-3 discussed npm packages are \texttt{react, typescript, and webpack}. Again, the top-5 libraries that are less frequently discussed in SO are \texttt{mocha, eslint, chai, babel-core, and lodash}. \syful{One possible reason is that, well-known libraries are well documented and may have their own forum, chat tools, etc. For this reason, there is no need to discuss them in SO.}
Furthermore, we find that 87.95\% of package co-usage mined from SO exist in the latest npm package release. 

\textbf{Lesson 2:} Developers post answers provided with usage example or execute command. Results do indicate the potential for a recommendation system, especially with the available execute commands and examples.
    
Although SO has been a useful resource for finding answers to questions, we find that popular and highly used libraries are not discussed as often.
However, we can see that the accepted answers may prove useful, as we believe that the usage examples and executable commands could be reused for tool support.

% The main contributions of this paper are summarized as follows:
% (i) a quantitative study on the library usage information mined from SO, and \texttt{npm} packages.
% (ii) a qualitative study on accepted answer posts of SO to uncover developers useful responses to solve library usage issues.
\syful{
The remainder of the paper is organized as follows. 
Section~\ref{sec:motivation} presents motivating example and research questions.
Section~\ref{sec:Dataset} describes the data preparation.
Section~\ref{sec:data_analysis} presents the analysis approach.
Section~\ref{sec:results} reports the results for each research question.
Section~\ref{sec:implication}, discusses the implications from this study.
Section~\ref{sec:relatedworks}, presents the related works and the research gap.
Section~\ref{sec:threats} discloses the threats to validity of our study. Finally, we conclude the paper in section~\ref{sec:conclusion} .}

\begin{figure}[t]
	\centering
	\includegraphics[width=0.49\textwidth]{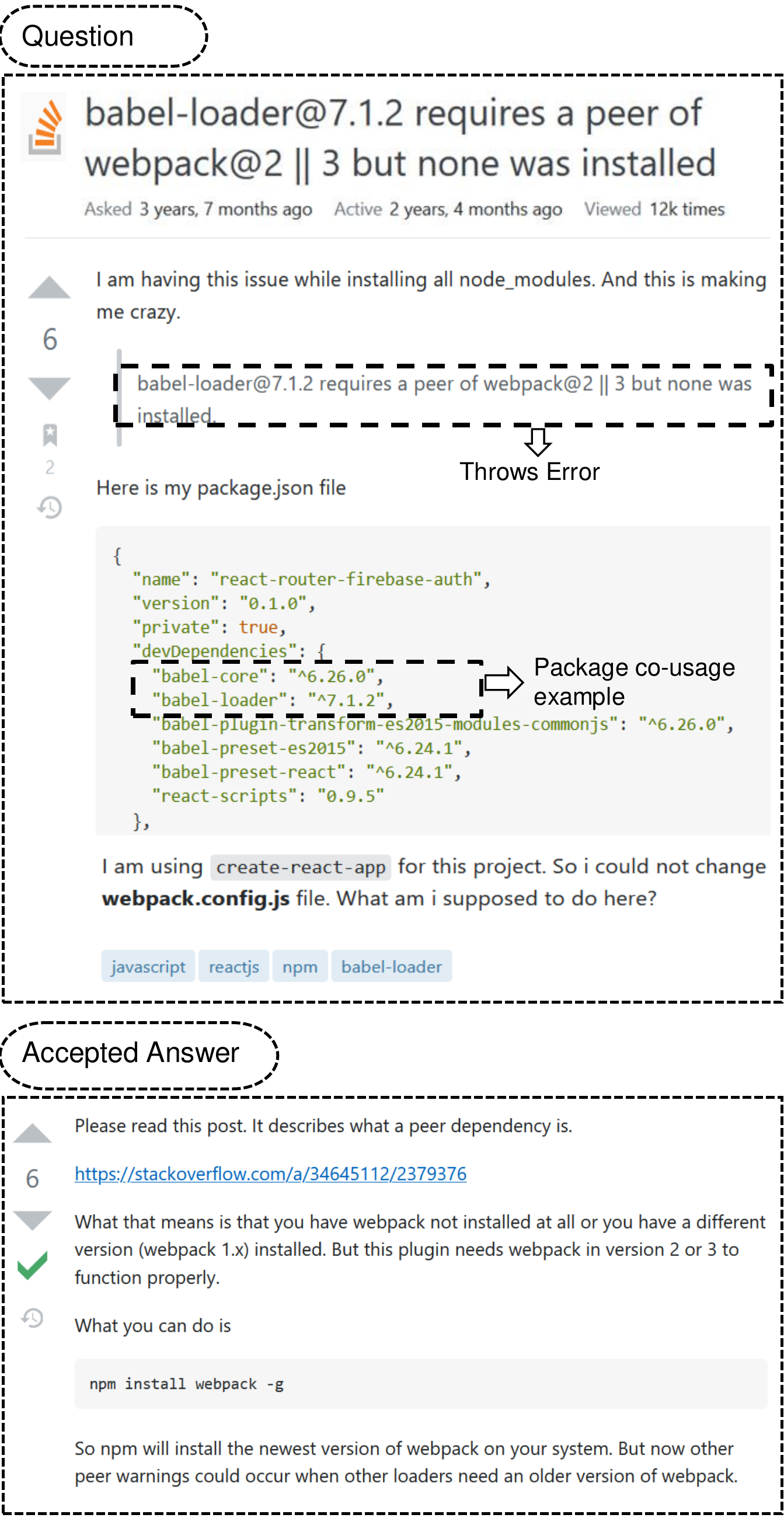}
	\caption{\syful{A motivating example of npm package co-usage in Stack Overflow. The example shows that a developer encounters a issue when installing all node\_modules, due to two dependent packages \texttt{babel-loader} and \texttt{webpack}.\footnote[2]{\url{https://stackoverflow.com/questions/46742824}}}}
	\label{fig:example}
\end{figure}

\begin{figure*}[t]
	\centering
	\includegraphics[width=0.8\textwidth]{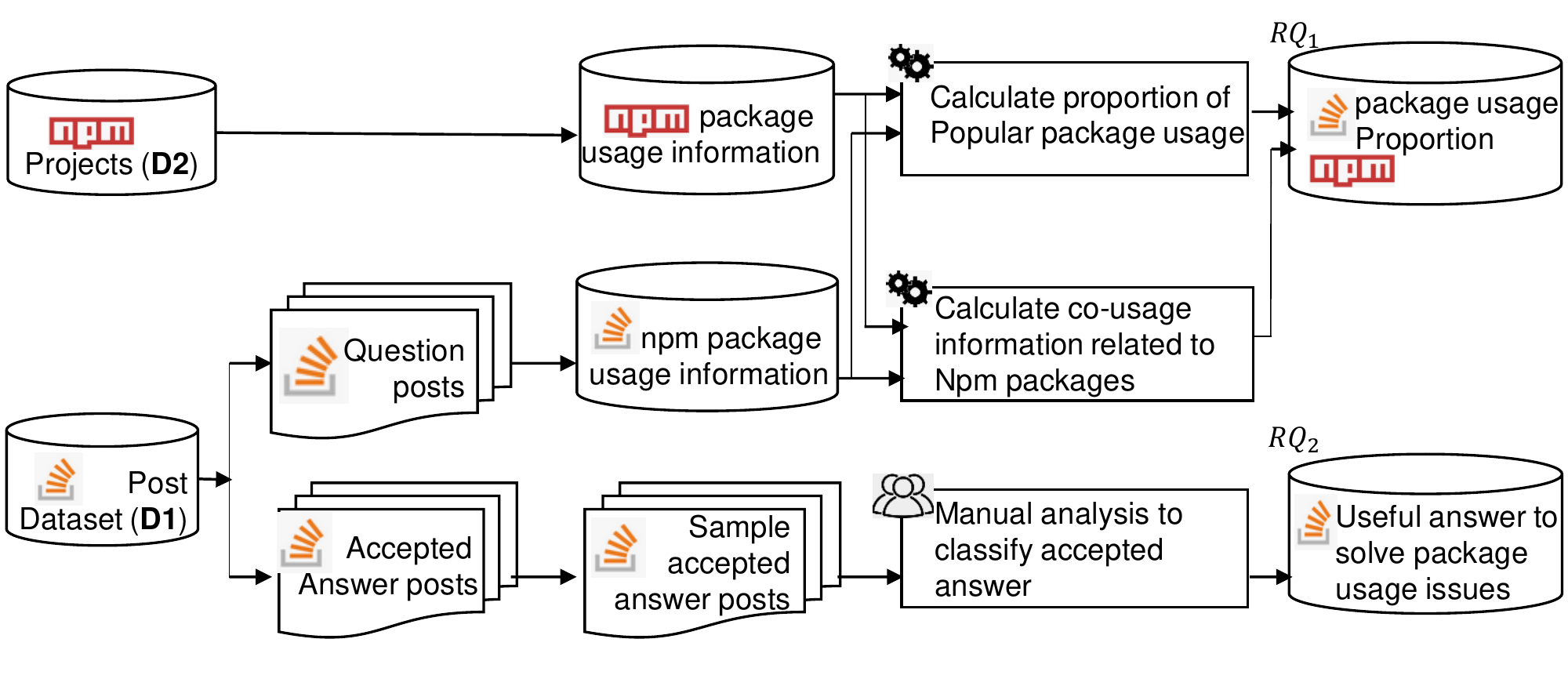}
	\caption{An overview of the methodology of our study.} 
	\label{fig:methodology}
\end{figure*} 
% \raula{where is the collection and where is the processing???}
\section{Motivating Example}
\label{sec:motivation}

Recent studies point out that SO is a useful question-answering site among developers to communicate various issues~\cite{chen2016similartech, mahajan2020recommending,treude2016augmenting,rubei2020postfinder, uddin2020mining}. In this paper, our motivation is to investigate the following assumptions:
\begin{itemize}
    \item Package usage information mined from SO is useful to solve developers issues while using libraries.
    \item Developer responses to package usage information in SO follow some useful patterns that might be reused by the recommendation tools.
\end{itemize}
Figure~\ref{fig:example} shows an example of a package co-usage related question post from SO.\footnote[2]{\url{https://stackoverflow.com/questions/46742824}}
\syful{As shown in the figure, a developer posts a question on the error issue of node\_module installation, resulting from two dependent packages \texttt{babel-loader} and \texttt{webpack}.
A closer look into the question description, we observe that the successful installation of \texttt{babel-loader@7.1.2} requires the package dependency of \texttt{webpack@2||3}. 
This issue is solved by a simple installation command (i.e., \texttt{npm install webpack -g})   mentioned in the accepted answer of the question.}
Such a motivating example suggests that package usage information mined from question answering sites may help solve package-related issues.
\\

\textit{\textbf{Research Questions:}} Our goal in this paper is to investigate whether package usage information mined from SO can help maintain the packages.
We formulate two research questions to guide our study.
\begin{itemize}
\item \textbf{\RQOne}\\
\textit{\uline{Motivation.}}
Developers often share package usage information to communicate various software development issues through SO.
We would like to understand what is the difference in the package usage information between SO and npm projects.

\item \textbf{\RqTwo}\\
\textit{\uline{Motivation.}}
This research question investigates the developer's response to package usage information discussed in SO posts. We argue that a closer look at the answers may reveal practical insights to improve real developers' experience dealing with package co-usage issues.

\end{itemize}

\section{Data Preparation}
\label{sec:Dataset}

\begin{table}[t]
\caption{Summary of dataset used in the study.}
\label{tab:dataset}
\resizebox{0.45\textwidth}{!}{%
\begin{tabular}{@{}lrr@{}}
\toprule
Data Source & Initial dataset & Final dataset \\ \midrule
D1: SO (npm question posts) & 30,136 & 2,100 \\
D2: libraries.io (npm projects) & 100,5955 & 217,934 \\ \bottomrule
\end{tabular}%
}
\end{table}

Our study exclusively examines the \texttt{npm} package usage information from Stack Overflow. 
Stack Overflow is the largest and most popular question-answering site among developers, which allows them to ask developers and experts for development related questions.
In addition, to compare with the package usage from the reality, we collect another dataset from the \texttt{libraries.io}.\footnote[3]{\url{https://libraries.io/}}
\texttt{libraries.io} is popularly known to monitor package releases.
Below, we describe the details of two studied datasets.

\textbf{\textit{(D1) from Stack Overflow posts}}: 
We rely on the SOTorrent~\cite{Baltes2018SoTorrent} to download the Stack Overflow data dump. 
We collect npm related question posts using a semi-automatic method.
To do so, we first extract all tags from the question posts, and then we manually check whether or not the tags are directly related to the npm packages.
After the examination, a list of eight tags that reflect npm packages posts are generated, i.e., `npm', `npm-install', `npm-script', `npm-ignore', `pnpm', `npm-shrinkwrap', `npm-start', `npm-build'.
We automatically identify all question posts using the defined tag list, resulting in 30,136 questions related to npm packages.

Next we further extract the npm related questions that contain the package usage information. We observed that several package names are as common as the natural language, i.e., \texttt{i}, \texttt{moment}, \texttt{should}, \texttt{express}, etc.). Thus, to reduce the bias, we only take those question posts that contain \texttt{package.json} files, resulting in 2,805 question posts.
As we focus on the relatively popular libraries, we extract all packages from these question posts and sort out 628 npm packages whose frequency are greater than ten.

To ensure that our sample dataset contains most npm libraries, we use the cumulative sampling method to assure that our question posts are saturated to cover all 628 npm packages.
Finally, our Stack Overflow npm package usage dataset consists of 2,100 question posts, as shown in Table~\ref{tab:dataset}.

\textbf{\textit{(D2) from npm packages}}: To compare the package usage information with SO ones, we construct a dataset consisting of npm packages from \texttt{libraries.io}. 
To do so, we first downloaded the latest history data dump, which is available at \texttt{https://libraries.io/data}, resulting in a total number of 1,005,955 \texttt{npm} project release history.

Similar to the \textbf{\textit{(D1)}}, we extract the libraries from these 1,005,955 projects and sort out 23,870 npm packages whose frequency are greater than ten.
In the end, our libraries.io npm package usage dataset consists of 217,934 npm projects using the cumulative sampling method, as shown in Table~\ref{tab:dataset}.

\section{Data Analysis}
\label{sec:data_analysis}
In this section, as shown in Fig. 2, we describe in detail the approaches
we use to answer two research questions.

\subsection{ Approach for RQ$_1$}
We perform an exploratory study to understand to which extents do developers discuss the package usage information from SO.
Below, we describe our approach in detail.

\textit{\textbf{Proportion of popular package usage:}}  To analyze the proportion of package usage in SO and npm projects, we extracted libraries from SO posts obtained in datasets D1 and D2, separately. Afterward, we count and sort these packages based on the frequency.

\textit{\textbf{Co-usage information related to npm package:}} To analyze the frequency of package usage information, we first need to identify npm package co-usage.
To do so, we follow the two steps: (I) we extract all the target packages appearing in the code snippets from 2,100 SO post related to npm package obtained in Section 3 (D1). 
\syful{Then we generate all possible binary combinations for co-usage of npm packages. For example, if a project contain three package dependencies ($A$, $B$, $C$), then the generated list of binary package co-usage will be: ($A$, $B$), ($A$, $C$), ($B$, $C$). 
After this step, we were able to retrieve 68,750 npm package co-usage information from SO.
(II) we then extract the npm package co-usage information based on the latest release, referring to 217,934 npm projects in Section 3 (D2). Finally, we check the occurrences of SO npm package co-usage information in the generated package co-usage from latest npm projects using the following formula: $\frac{\alpha}{\beta}\times100$ where $\alpha$=Number of SO npm package co-usage found in the latest npm project release, and $\beta$=Total npm package co-usage extracted from SO. }
% Thus we find that 87.95\% (i.e., $\frac{60,470}{68,750}\times100$) of the SO package co-usage information exists in the latest npm project release.
% Afterward, to calculate the frequency of each co-used package in the latest packages, we use the following formula: $\frac{Users\ of\ both\ A\ and\ B}{Users\ of\ A}$ and $\frac{Users\ of\ both\ A\ and\ B}{Users\ of\ B}$, where ($A$, $B$) is a package co-usage extracted from SO.

% % ($\frac{60,470}{68,750}\times100$)  
    
In addition, to understand the issues raised by package related question,
we extract the title information from 2,100 Stack Overflow posts obtained in D1 and automatically extracted the keywords using traditional Nature Language Processing (NLP) , including stop word removals. 
The output is a corpus of title keywords.

\begin{figure}[t]
	\centering
	\includegraphics[width=0.45\textwidth]{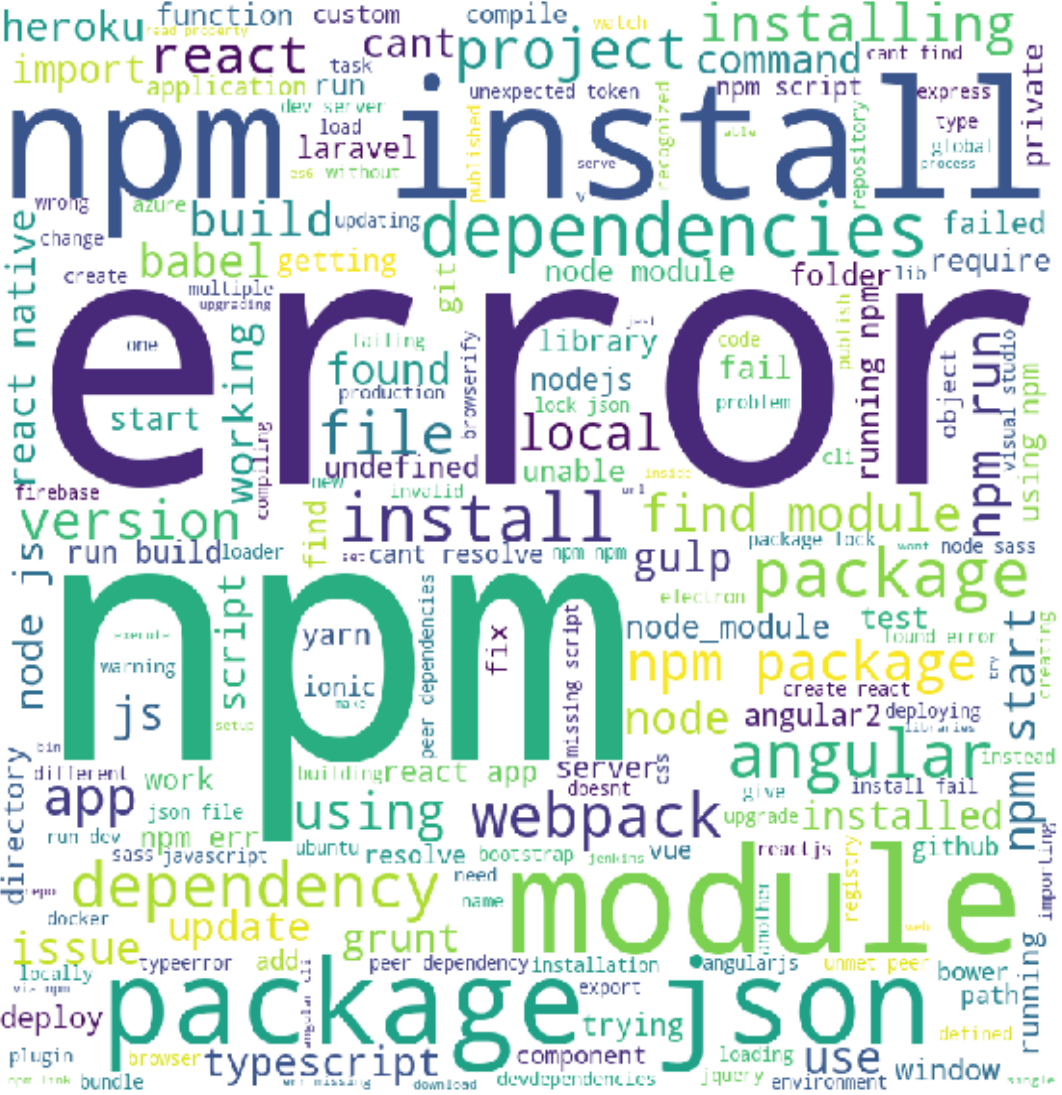}
	\caption{\syful{Word cloud generated from SO npm posts title that contains package usage information. The word cloud shows that npm posts are primarily related to various types of errors.}} 
	\label{fig:RQ1_cloud}
\end{figure}

To visually understand the package related issues asked in the SO, we generate a word cloud based on the constructed title corpus.

\textit{\textbf{Observation 1- The npm post that discuss  package  usage information are mostly relate to different type of errors.}}
Figure.~\ref{fig:RQ1_cloud} shows the Word cloud based on SO posts titles. The word cloud shows that npm posts regarding  package usage information are primarily related to various types of errors like installation error, build failure, etc.
% This result hints that the application developers tried to avoid SO  package co-usage in npm projects due to various incompatibility issues. Therefore, 
% this information's can be re-used to keep developers aware during software development.

\subsection{ Approach for RQ$_2$}

We conduct a qualitative analysis to investigate the accepted answer post from SO. 
We analyze the accepted answer since these answers are solutions that work for developers.\footnote[4]{\url{https://stackoverflow.com/tour}}
Below we describe the representative sample construction and the manual coding process.

\textit{\textbf{Representative sample construction:}} As the full set of our constructed data is too large to manually examine their accepted answers, we then draw a statistically representative sample.
% \raula{wait I saw this in the Fig 2 and came out here.... is this data collection again, or what?}
The representative sample consists of 286 randomly selected accepted answer, with a confidence level of 95\% and a interval of 5.~\footnote[5]{\url{https://www.surveysystem.com/sscalc.htm}} 

\textit{\textbf{Manual coding:}} We adopt three rounds to do our manual coding.
First, the three authors independently sampled 25 random questions and conducted an open discussion to establish an initial code taxonomy.
Hence, the following three codes emerged from our manual analysis in the first round.

\begin{figure}[t]
	\centering
	\includegraphics[width=0.48\textwidth]{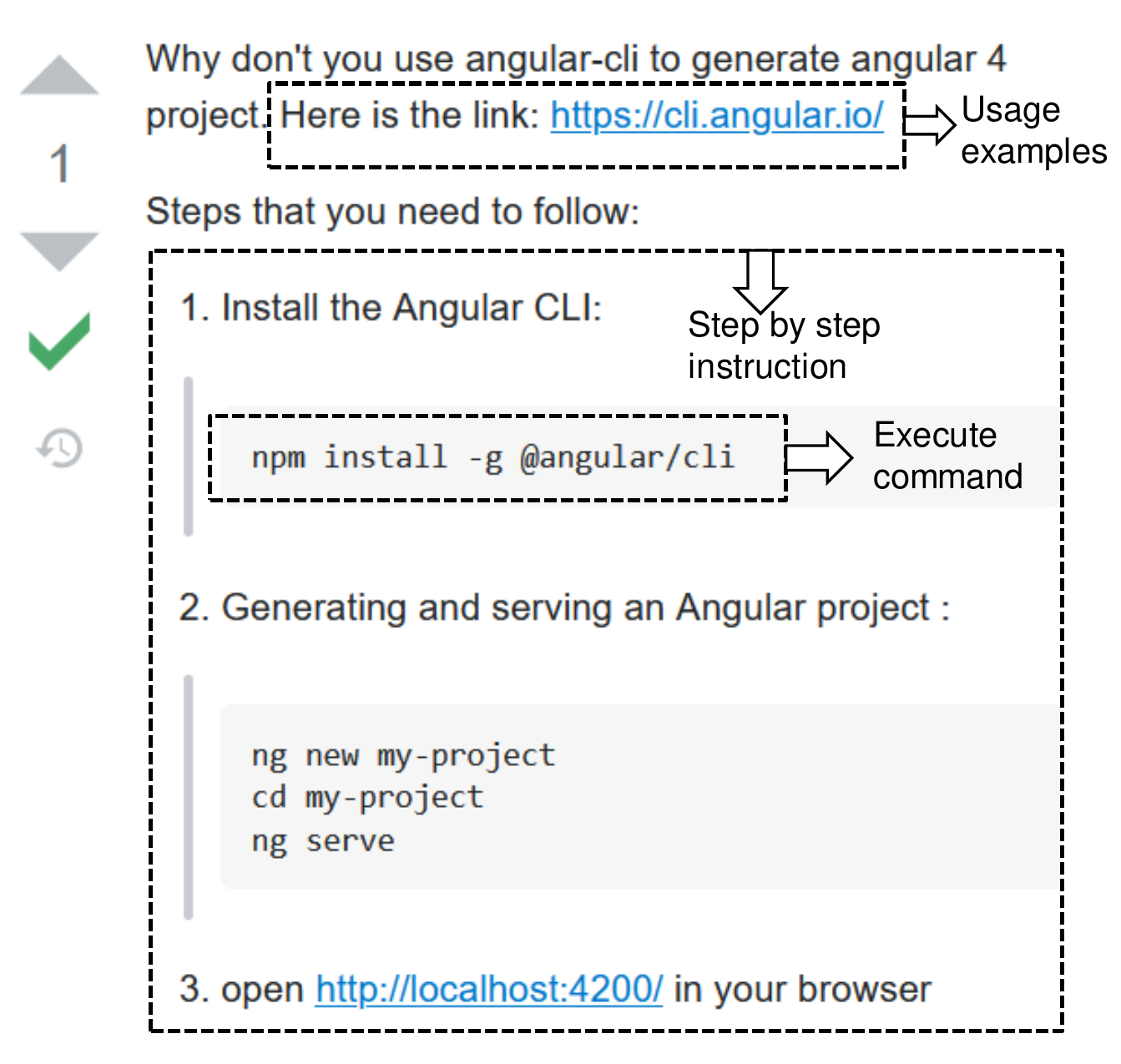}
	\caption{An example that motivates to classify developers response. In the answer we observe that, it contains usage examples, execute command, and step by step instruction.
	} 
	\label{fig:example2}
\end{figure}

% \raula{it would be nice to put an example and show so that the reader can understand}
\begin{itemize}
     \item \textit{Execute command}: The accepted answer explicitly mentions executing commands. In definition, execute commands describe the process of running a computer software program, script, or command. 
     As shown in Fig.~\ref{fig:example2}, \texttt{npm install -g @angular/cli} is an execute command to install `angular/cli'.
 
\item \textit{Step by step Instruction}: The accepted answer contains step by step information to get the solution. In definition, instructions are detailed (i.e., step by step) information about how something should be done or operated. 
\syful{As shown in Fig.~\ref{fig:example2}, the accepted answer contains three distinct step to solve the library usage issue (i.e., Angular CLI installation, generating and serving an Angular project, and open local host page in the browser.)}
 
\item \textit{Usage Example}: \syful{The accepted answer explicitly mentions examples, referred to external links, source code, configuration files, etc.} In definition, usage examples are defined as models, or typical cases (like external links, source code, etc.) used to solve a problem. 
As shown in Fig.~\ref{fig:example2}, the external link mention in the beginning of the answer is usage example.
%  \raula{I think these are very easy to tell the difference and we had some keywords.}
 \end{itemize}
 In the second round, to assure that there is no new cases, the three authors coded another 25 samples and we found that the initialized codes can totally fit.  
In the third round, to test the comprehension understanding, we coded another 20 samples and used the Kappa score to measure the agreement.
The score is 0.82, which is implied as nearly perfect~\citep{viera2005understanding}.
Based on this encouraging result,
the first author then coded the rest of the representative sample independently.
% The three distinct response by developers for library usage are explained below:
 
\section{Results}
\label{sec:results}
In this section, we provide the results for each research question. First, we describe the result analysis, and then highlight our findings and answer
each question.

\subsection{Answering RQ$_1$}
To show the proportion of popular package usage, we depict tables to statistically show the package usage between SO and npm projects. \syful{Then, to analyze the frequency of So npm package co-usage in the latest npm projects, we calculated the ratio using formula (i.e., $\frac{\alpha}{\beta}\times100$ ) discussed in the approach.}
% \syful{Then}, to show the frequency of co-usage libraries \syful{in the latest npm projects} that are extracted from SO posts,  we use a box plot and a histogram plot. 

% \begin{figure}[t]
% 	\centering
% 	\includegraphics[width=0.36\textwidth]{Fig/boxplot_so_cousage.pdf}
% 	\caption{\syful{Frequency of SO package co-usage in latest npm projects. Analysis result indicate that package co-usage extracted from SO also exist in the latest npm projects but with low frequency (i.e., median frequency 1.2\%).}} 
% 	\label{fig:RQ1_Boxplot}
% \end{figure}
% \textcolor{red}{I cannot understand this figure. X-axis and Y-axis may be renamed.}
\begin{table}[t]
\caption{Top-15 npm packages extracted from SO posts with their proportion and rank in the latest npm projects. Result shows that Only three out of top-10 npm packages are mostly discussed in SO.}
\label{tab:libusage}
\resizebox{0.49\textwidth}{!}{%
\begin{tabular}{@{}lrrrr@{}}
\toprule
npm packages & \begin{tabular}[c]{@{}r@{}}Count\\ (SO)\end{tabular} & \begin{tabular}[c]{@{}r@{}}Count \\ (npm \\ projects)\end{tabular} & \begin{tabular}[c]{@{}l@{}}Rank \\ (SO)\end{tabular} & \begin{tabular}[c]{@{}r@{}}Rank\\  (npm \\ projects)\end{tabular} \\ \midrule
react & 586 & 42, 591 & 1 & 9 \\
typescript & 548 & 57,864 & 2 & 4 \\
webpack & 489 & 52,453 & 3 & 5 \\
rxjs & 471 & 12,339 & 4 & 69 \\
zone.js & 462 & 8,570 & 5 & 119 \\
react-dom & 461 & 32,941 & 6 & 16 \\
@angular/core & 434 & 10,050 & 7 & 95 \\
@angular/common & 434 & 9,406 & 8 & 103 \\
@angular/compiler & 426 & 9,170 & 9 & 110 \\
@angular/platform-browser & 424 & 8,482 & 10 & 120 \\
@angular/platform-browser-dynamic & 419 & 7,634 & 11 & 132 \\
jquery & 413 & 9,263 & 12 & 108 \\
@angular/forms & 401 & 6,608 & 13 & 147 \\
@angular/http & 388 & 5,433 & 14 & 169 \\
@angular/router & 380 & 5,583 & 15 & 166 \\ \bottomrule
\end{tabular}%
}
\end{table}

\begin{table}[t]
\centering
\caption{Top-15 package usage extracted from the latest npm projects with their proportion and rank in the SO posts. The top package usage patterns from npm projects shows that application developers top usage packages are different from SO.}
\label{tab:libusage1}
\resizebox{0.49\textwidth}{!}{%
\begin{tabular}{@{}lrrrr@{}}
\toprule
npm package & \begin{tabular}[c]{@{}r@{}}Count\\ (npm projects)\end{tabular} & \begin{tabular}[c]{@{}r@{}}Count\\ (SO)\end{tabular} & \begin{tabular}[c]{@{}r@{}}Rank \\ (npm projects)\end{tabular} & \begin{tabular}[c]{@{}r@{}}Rank\\ (SO)\end{tabular} \\ \midrule
mocha & 101898 & 126 & 1 & 65 \\
eslint & 81767 & 199 & 2 & 45 \\
chai & 58368 & 114 & 3 & 78 \\
\rowcolor[HTML]{C0C0C0} typescript & 57864 & 548 & 4 & 2 \\
\rowcolor[HTML]{C0C0C0} webpack & 52453 & 489 & 5 & 3 \\
babel-core & 51351 & 309 & 6 & 26 \\
lodash & 45618 & 348 & 7 & 19 \\
babel-loader & 44398 & 340 & 8 & 22 \\
\rowcolor[HTML]{C0C0C0} react & 42591 & 586 & 9 & 1 \\
jest & 41188 & 115 & 10 & 76 \\
babel-eslint & 39336 & 134 & 11 & 68 \\
babel-cli & 38038 & 102 & 12 & 83 \\
eslint-plugin-import & 36019 & 94 & 13 & 90 \\
@types/node & 35197 & 320 & 14 & 24 \\
rimraf & 34466 & 139 & 15 & 63 \\ \bottomrule
\end{tabular}%
}
\end{table}

	\begin{table}[t]
	\centering
\caption{\syful{Top-15 package co-usage extracted from SO posts except \texttt{angular} since rest of the top co-usage are related to \texttt{angular}. The top co-usage patterns and their rank in SO indicate that developers discuss most package dependency issues related to \texttt{angular} followed by \texttt{(`typescript',   `zone.js')}.}}
\label{tab:cousage}
\resizebox{0.48\textwidth}{!}{%
\begin{tabular}{@{}lcc@{}}
\toprule
Package Co-usage & Rank  & Count \\ \midrule
(`typescript',   `zone.js') & 9 & 317 \\
(`zone.js', `rxjs') & 15 & 290 \\
(`react-dom',   `react') & 17 & 288 \\
(`typescript',   `rxjs') & 21 & 283 \\
(`karma',   `karma-jasmine') & 31 & 258 \\
(`zone.js',   `core-js') & 33 & 251 \\
(`core-js', `rxjs') & 39 & 233 \\
(`webpack',   `babel-loader') & 40 & 230 \\
(`typescript',   `core-js') & 40 & 230 \\
(`jasmine-core',   `karma-jasmine') & 43 & 225 \\
(`karma-jasmine',   `karma-chrome-launcher') & 44 & 223 \\
(`babel-core', `babel-loader') & 45 & 220 \\
(`typescript', `tslint') & 46 & 216 \\
(`typescript', `karma') & 47 & 210 \\
(`typescript', `karma-jasmine') & 48 & 209 \\ \bottomrule
\end{tabular}%
}
\end{table}

\textbf{\textit{ \syful{Observation 2-} Only three out of top-10 npm packages are mostly discussed in SO.}} Table~\ref{tab:libusage} shows the top-15 packages discussed in SO with their proportion and ranks in the latest npm projects. The top-3 discussed npm packages are \texttt{react}, \texttt{typescript}, and \texttt{webpack}. Again, Table~\ref{tab:libusage1} shows the top-15 package usage extracted from the latest npm projects. We observe that, the top-5 packages in the latest npm projects which are less frequently discussed in SO are \texttt{mocha, eslint, chai, babel-core, and lodash.}  \syful{One possible reason is that, such well-known libraries are well documented and may have their own forum, chat tools, etc. For this reason, there is no need to discuss them in SO.}

% \textbf{\textit{ Observation 3- npm package usage information can be detected from Stack Overflow.  }}

\textbf{\textit{Observation 3- 87.95\% of the SO package co-usage information exist in the latest npm project release.}} \syful{ Furthermore, Table~\ref{tab:cousage} shows the \syful{top-15} SO package co-usage mentioned by developers. We observed that most of the package co-usage mentioned by developers are related to \texttt{angular} followed by \texttt{(`typescript',   `zone.js')}}. The top co-usage patterns from SO and their rank hints that developers face most error type issues when they use angular packages. 

% Figure~\ref{fig:RQ1_Boxplot} shows the SO npm package co-usage frequency in the latest npm project release. We observe that package co-usage extracted from SO also exist in the latest npm project release but with low frequency. The median of package co-usage frequency is 1.2\%.
% As shown in Fig.~\ref{fig:RQ1} most of the package co-usage mined from SO are likely to occur with a low-frequency range between 0.29-4.39\% in the latest \texttt{npm} project release. We also observed that, as the frequency increased, the number of package co-usage decreased.  

\begin{tcolorbox}
    \textbf{RQ$_1$ Summary}: 
 Our analysis result shows that,  only three out of top-10 npm packages are mostly discussed in SO. In addition, 87.95\% of the SO npm package co-usage information exist in the latest npm project release.
%  \raula{your question ask for proportion, yet your analysis shows frequency?}
\end{tcolorbox}

\subsection{Answering RQ$_2$}
\syful{
To show the useful accepted answer attributes pattern in response to the package usage question by developers, we prepare all possible combinations for three manually curated attributes: Execute command, Step by step instruction, and Usage example, respectively. Thus, we obtain eight distinct combinations (i.e., subsets), including the others (i.e., empty set). Finally, we calculate the percentage of each variety in the manually analyzed representative sample.}

\begin{figure}[t]
	\centering
	\includegraphics[width=0.48\textwidth]{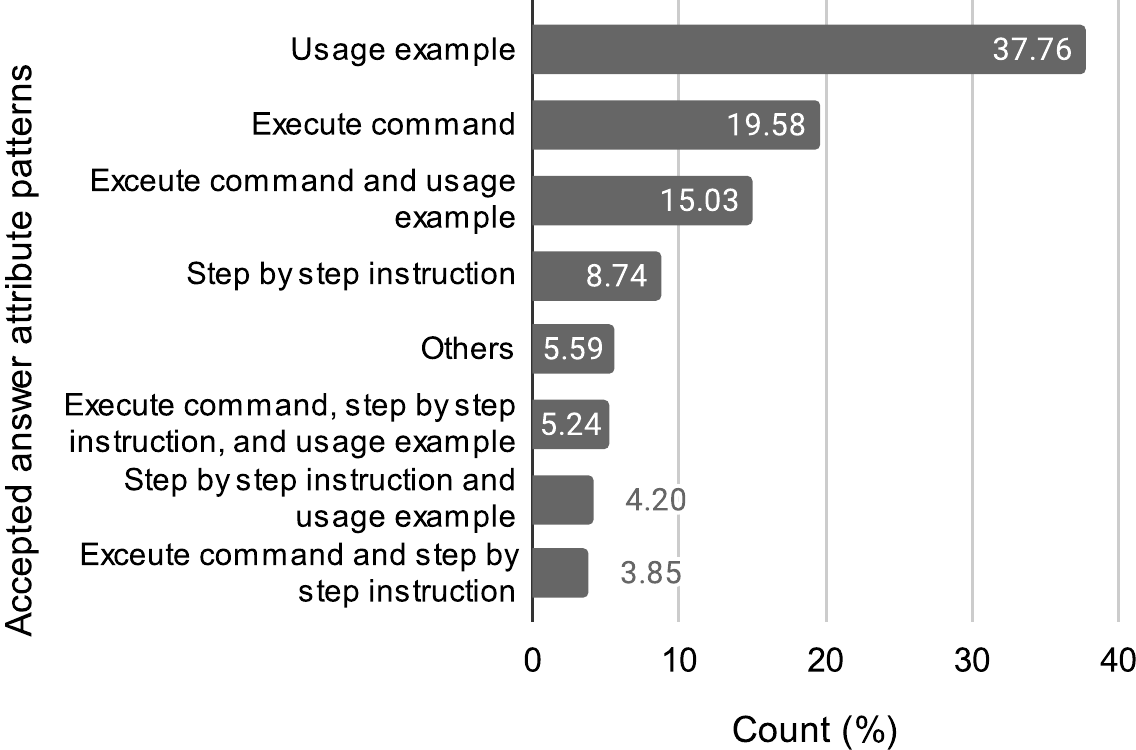}
	\caption{\syful{Analysis of accepted answers posted in response to questions that include package usage information. Result shows that 37.76\% accepted answers contain \texttt{usage example} followed by \texttt{execute command} 19.58\%.}} 
	\label{fig:RQ2_1}
\end{figure}

\textbf{\textit{Observation 4- Our results show that, accepted answers posted by developers in response to questions that include package usage information mostly contain usage examples (i.e., includes real-life examples, external links, source code, build configuration files, etc.).}} Figure~\ref{fig:RQ2_1} shows the analysis result of accepted answers posted by developers in response to questions that include package usage information. We observe that \texttt{usage example} (36.76\%) is most dominant attribute in accepted answer, followed by \texttt{execute command} (19.58\%). The third most frequent (15.03\%) attribute combination in the accepted answer is \texttt{execute command and usage example}.
These findings hint that application developers are suffering from a lack of technical depth in managing third-party libraries in their applications.\\

\begin{tcolorbox}
    \textbf{RQ$_2$ Summary}: 
   Result shows that 37.76\% accepted answers posted by developers in response to questions that include package usage information contain \texttt{usage example} followed by \texttt{execute command} 19.58\%. 
\end{tcolorbox}

\section{Discussion}
\label{sec:implication}
In order to aid application developers faced with package usage issues, we conducted an empirical study to understand the usefulness of package usage information mined from SO. We learned two lessons along the way:

\textbf{\textit{Lesson 1-}}
Many of the library usage information on SO is not from the popular npm package. 
We find that only three out of the top ten of the most used npm libraries are mentioned in SO. The top-3 discussed npm packages are \texttt{react, typescript, and webpack}. Again, the top-5 libraries that are less frequently discussed in SO are \texttt{mocha, eslint, chai, babel-core, and lodash}.
Furthermore, we find that 87.95\% package co-usage mined from SO exist in the latest npm package release. The npm post that discuss package usage information are mostly relate to different type of errors. 
    
\textbf{\textit{Lesson 2-}} Developers tend to post answers that are usage example or execute command. 
    The good news is that maybe the answers can be useful for any npm developer that suffers from similar issues. 
    There is potential for a recommendation system, especially with the available execute commands and examples available.

\section{Related Work}
\label{sec:relatedworks}
In this section, we discuss the related works. First, we discuss on third-party package usage issues faced by developers. Second, we discuss on mining useful information from question-answering sites.\\

\textbf{\textit{Third-party libraries usage issues:}}
software package is a reusable component developed by a body other than the original vendor of the development platform.
 The usage of third-party libraries provides developers with unique opportunity to integrate pre-tested, reusable software that saves development time and cost.\footnote[6]{\url{https://tinyurl.com/y5b2fajq}} Recent empirical studies have found that 93.30\%
of modern OSS project use third-party libraries, with an average of 28 libraries per project~\cite{thung2013automated}. 

In recent years, analyzing the characteristics of software ecosystem has gained much attention. Decan et al.~\cite{decan2019empirical} investigated package dependency network for seven software ecosystems. Their findings reveals that, software ecosystems grew over time in term of number of published libraries. Bogart et al.~\cite{bogart2016break} investigate the challenges of reusing libraries from software ecosystem. They reported that developers struggle with changing versions of the libraries as the changes might potentially break dependent codes. Bavota et al.~\cite{bavota2015apache} examine the evolution of dependencies in Apache ecosystem  and found that developers were reluctant to upgrade their dependencies considering that changes of a package might break its dependent libraries. Xavier et al.~\cite{xavier2017historical} performed a large scale study on 317 real-life Java libraries with 9K releases and 260K projects. Their analysis results show that 14.78\% of API changes are incompatible with previous versions. Kula et al.~\cite{kula2018developers} also reported that, developers do not update their dependencies even though the updates are related to new features, fix vulnerabilities. Several research was done on package recommendation tools like LibRec~\cite{thung2013automated}, LibCup~\cite{ouni2017search}, CrossRec~\cite{nguyen2020crossrec}. In this research, we empirically investigate usefulness of npm package usage information mined from question-answering site.\\

\textbf{\textit{Mining SO:}}
Recent studies point out that SO is a useful source for developers to meet their information needs. 
For instance, Chen et al.~\cite{chen2016similartech} reported that SO is useful in recommending analogical libraries across different programming languages. Mahajan et al.~\cite{mahajan2020recommending} proposed a recommendation tool to fix Run-time Exceptions based on knowledge from SO posts. 
Similarly, Treude et al.~\cite{treude2016augmenting}, Rubei et al.~\cite{rubei2020postfinder}, and Uddin et al.~\cite{uddin2020mining} showed that SO posts are useful knowledge source to support software developers. 
Previous studies suggest that SO can be useful to solve package usage related issues.  
There is no existing work that studies package usage information mined from SO help to improve developers' experience.

\section{\syful{Threats to validity}} 
\label{sec:threats}
In this section, we discuss threats to validity that might influence our study.\\

\textbf{\textit{Internal Validity:}}  Threats to internal validity refer to experimental bias. In this study, we found two main internal threads that could affect our results. First, is the pre-processing of the dataset we decide the number of posts (2100) and npm packages (217,934) based on cumulative extraction of npm libraries and the generated co-usages. We continue the cumulative extraction until all the libraries and the co-usage cover. Second, in RQ$_2$ we perform manual analysis on random sample since the dataset size is large. To mitigate this challenge, we prepare representative sample consists of 286 randomly selected accepted answer, with a confidence level of 95\% and a interval of 5.\\
% ~\footnote[4]{\url{https://www.surveysystem.com/sscalc.htm}} 

\textbf{\textit{External validity:}} Threats to external validity refer to the generalizability of our findings. Our datasets consist of npm packages from libraries.io and SO posts. SO is a popular platform for question and answers from developers with various domains and experts. Hence, our observations and results can not be generalized for other package managers like \texttt{\syful{Maven}, NuGet,} and others. Besides, we consider only those SO posts that contain \texttt{package.json} file. Selecting more question posts may cause variation of top package co-usage results. \\

\textbf{\textit{Construct validity:}} \syful{Threats to construct validity refers to the suitability of our evaluation measure. 
In our qualitative analysis of classifying accepted answers (RQ2), the answer patterns may be miscoded due to the subjective nature of our coding approach. To mitigate this threat, we took a systematic approach to validate the taxonomy and the comprehension understanding by the three authors in several rounds. Only until the Kappa score reaches 0.82, indicating that the agreement is almost perfect (0.81-1.00), we were able to complete the rest of the sample dataset.}

\section{Conclusion}
\label{sec:conclusion}
In this paper, we examine the usefulness of package usage information mined from SO. We perform a case study on npm package co-usage information from SO question posts (2100) and libraries.io (217,934 npm projects) dataset. Although SO has been a useful resource for finding answers to questions, we find that unfortunately popular and highly used libraries are not discussed as often.
However, we can see that the accepted answers may prove useful, as we believe that the usage examples and executable commands could be reused or be used for tool support. In our future work, we will develop tool support that will utilize SO usage examples and executable commands extracted from accepted answers to assist npm application developers. 

\section*{Acknowledgment}
This work has been supported by JSPS KAKENHI Grant Numbers JP8H04094, JP20K19774, and JP20H05706.

\bibliographystyle{ieicetr}
% \bibliography{reference}
%\bibliographystyle{ieicetr}% bib style
%\bibliography{}% your bib database

\begin{thebibliography}{99}% more than 9 --> 



\bibitem{dietrich2014broken}
J. Dietrich, K. Jezek, and P. Brada, “Broken promises: An empirical study into evolution problems in java programs caused by library upgrades,” 2014 Software Evolution Week-IEEE Conference on Software Maintenance, Reengineering, and Reverse Engineering (CSMR-WCRE), pp.64–73, IEEE, 2014.


\bibitem{raemaekers2017semantic}
S. Raemaekers, A. van Deursen, and J. Visser, “Semantic versioning and impact of breaking changes in the maven repository,” Journal of Systems and Software, vol.129, pp.140–158, 2017.

\bibitem{raemaekers2014semantic}
S. Raemaekers, A. Van Deursen, and J. Visser, “Semantic versioning versus breaking changes: A study of the maven repository,” 2014 IEEE 14th International Working Conference on Source Code Analysis and Manipulation, pp.215–224, IEEE, 2014.

\bibitem{de2018library}
F.L. De La Mora and S. Nadi, “Which library should i use?: a metric-based comparison of software libraries,” 2018 IEEE/ACM 40th International Conference on Software Engineering: New Ideas and Emerging Technologies Results (ICSE- NIER), pp.37–40, IEEE, 2018.

\bibitem{dietrich2019dependency}
J. Dietrich, D. Pearce, J. Stringer, A. Tahir, and K. Blincoe, “Dependency versioning in the wild,” 2019 IEEE/ACM 16th International Conference on Mining Software Repositories (MSR), pp.349– 359, IEEE, 2019.

\bibitem{thung2013automated}
F. Thung, D. Lo, and J. Lawall, “Automated library recommendation,” 2013 20th Working conference on reverse engineering (WCRE), pp.182– 191, IEEE, 2013.

\bibitem{ouni2017search}
A. Ouni, R.G. Kula, M. Kessentini, T. Ishio, D.M. German, and K. Inoue, “Search-based software library recommendation using multi-objective optimization,” Information and Software Technology, vol.83, pp.55–75, 2017.

\bibitem{saied2018improving}
M.A. Saied, A. Ouni, H. Sahraoui, R.G. Kula, K.	Inoue, and D. Lo, “Improving reusability of software libraries through usage pattern mining,” Journal of Systems and Software, vol.145, pp.164– 179, 2018.

\bibitem{alrubaye2020learning}
H. Alrubaye, M.W. Mkaouer, I. Khokhlov, L.	Reznik, A. Ouni, and J. Mcgoff, “Learning to recommend third-party library migration opportunities at the api level,” Applied Soft Computing, vol.90, p.106140, 2020.

\bibitem{yu2017combining}
H. Yu, X. Xia, X. Zhao, and W. Qiu, “Combining collaborative filtering and topic modeling for more accurate android mobile app library recommendation,” Proceedings of the 9th Asia-Pacific Symposium on Internetware, pp.1–6, 2017.


\bibitem{nguyen2020crossrec}
P.T. Nguyen,  J. Di Rocco,  D. Di Ruscio,  and M. Di Penta, “Crossrec: Supporting software developers by recommending third-party libraries,” Journal of Systems and Software, vol.161, p.110460, 2020.

\bibitem{wu2019developers}
Y. Wu, S. Wang, C.P. Bezemer,  and  K. Inoue, “How do developers utilize source code from stack overflow?,” Empirical Software Engineering, vol.24, no.2, pp.637–673, 2019.


\bibitem{chen2016similartech}
C. Chen and Z. Xing, “Similartech: automatically recommend analogical libraries across different programming languages,” Proceedings of the 31st IEEE/ACM International Conference on Automated Software Engineering, pp.834–839, 2016.


\bibitem{mahajan2020recommending}
S. Mahajan, N. Abolhassani, and M.R. Prasad, “Recommending stack overflow posts for fixing runtime exceptions using failure scenario matching,” Proceedings of the 28th ACM Joint Meeting on European Software Engineering Conference and Symposium on the Foundations of Software Engineering, pp.1052–1064, 2020.

\bibitem{treude2016augmenting}
C. Treude and M.P. Robillard, “Augmenting api documentation with insights from stack overflow,” 2016 IEEE/ACM 38th International Conference on Software Engineering (ICSE), pp.392–403, IEEE, 2016.

\bibitem{uddin2020mining}
G. Uddin, F. Khomh, and C.K. Roy, “Mining api usage scenarios from stack overflow,” Information and Software Technology, vol.122, p.106277, 2020.


\bibitem{10.1145/3368089.3409711}
E. Larios Vargas, M. Aniche, C. Treude, M. Bruntink, and G. Gousios, “Selecting Third-Party Libraries: The Practitioners’ Perspective’ perspective,” Proceedings of the 28th ACM Joint Meeting on European Software Engineering Conference and Symposium on the Foundations of Software Engineering, ESEC/FSE 2020, New York, NY, USA, p.245–256, Association for Computing Machinery, 2020.


\bibitem{xavier2017we}
L. Xavier, A. Hora, and M.T. Valente, “Why do we break apis? first answers from developers,” 2017 IEEE 24th International Conference on Software Analysis, Evolution and Reengineering (SANER), pp.392–396, IEEE, 2017.


\bibitem{todorov2017sol}
B. Todorov, R.G. Kula, T. Ishio, and K. Inoue,“Sol mantra: Visualizing update opportunities based on library coexistence,” 2017 IEEE Working Conference on Software Visualization (VISSOFT),pp.129–133, IEEE, 2017.


\bibitem{rubei2020postfinder}
R. Rubei, C. Di Sipio, P.T. Nguyen, J. Di Rocco,and D. Di Ruscio, “PostFinder: Mining Stack Overflow posts to support software developers,”Information and Software Technology,  vol.127,p.106367, 2020.


\bibitem{Baltes2018SoTorrent}
S. Baltes, L. Dumani, C. Treude, and S. Diehl, “Sotorrent: reconstructing and analyzing the evolution of stack overflow posts,” Proceedings of the 15th International Conference on Mining Software Repositories (MSR) 2018, ed. A. Zaidman,
Y. Kamei, and E. Hill, pp.319–330, ACM, 2018.

\bibitem{viera2005understanding}
A.J. Viera, J.M. Garrett, et al.,  “Understanding interobserver agreement: the kappa statistic,” Fam med, vol.37, no.5, pp.360–363, 2005.


\bibitem{decan2019empirical}
A. Decan, T. Mens, and P. Grosjean, “An empirical comparison of dependency network evolution in seven software packaging ecosystems,” Empirical Software Engineering, vol.24, no.1, pp.381–416, 2019.


\bibitem{bogart2016break}
C. Bogart, C. Kästner, J. Herbsleb, and F. Thung, “How to break an api: cost negotiation and community values in three software ecosystems,” Pro- ceedings of the 2016 24th ACM SIGSOFT International Symposium on Foundations of Software Engineering, pp.109–120, 2016.


\bibitem{bavota2015apache}
G. Bavota, G. Canfora, M. Di Penta, R. Oliveto, and S. Panichella, “How the apache community upgrades dependencies: an evolutionary study,” Empirical Software Engineering, vol.20, no.5, pp.1275–1317, 2015.


\bibitem{xavier2017historical}
L. Xavier, A. Brito, A. Hora, and M.T. Valente, “Historical and impact analysis of api breaking changes: A large-scale study,” 2017 IEEE 24th International Conference on Software Analysis, Evolution and Reengineering (SANER), pp.138–147, IEEE, 2017.

\bibitem{kula2018developers}
R.G. Kula, D.M. German, A. Ouni,  T.  Ishio, and K. Inoue, “Do developers update their library dependencies?,” Empirical Software Engineering, vol.23, no.1, pp.384–417, 2018.




\end{thebibliography}

%\profile{}{}
%\profile*{}{}% without picture of author's face

% \balance

 \profile[Authors_Photo/Syful.JPG]{Syful Islam}{He received the M.E. degree in Information Science from Nara Institute of Science and Technology, Japan. He is currently working toward the P.hD degree in the same institute. At present, he is on study leave from Noakhali Science and Technology University, Bangladesh. His research interests include software ecosystem, mining Stack Overflow, etc.}
\profile[Authors_Photo/WANG.JPG]{Dong Wang}{He received the M.E. degree in Information Science from Nara Institute of Science and Technology, Japan. He
is currently working toward the Doctor degree in Nara Institute of Science and Technology, Japan. His research interests include code review and mining software repositories.}
\profile[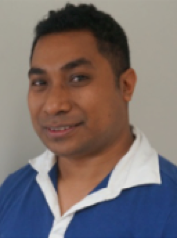]{Raula Gaikovina Kula}{is currently an assistant professor at Nara Institute of Science and technology. In 2013, he graduated with a PhD. from Nara Institute of Science and Technology, Japan. He is currently an active member of the IEEE Computer Society and ACM. His research interests include repository mining, code review, software libraries and visualizations.}

\profile[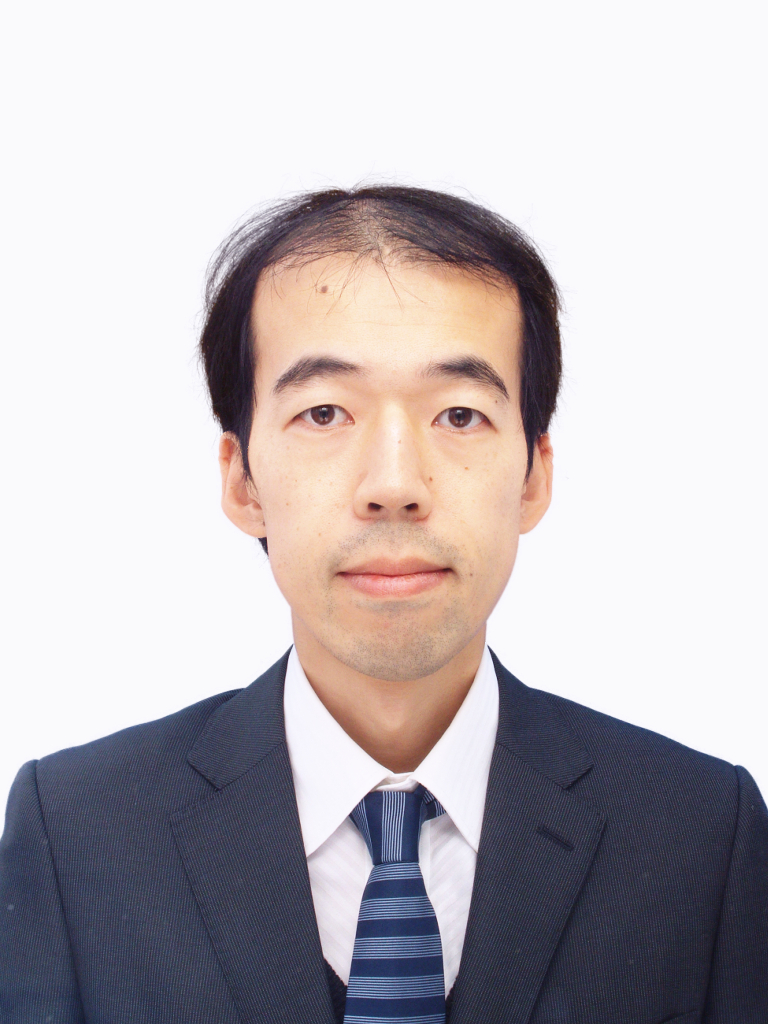]{Takashi Ishio}{received the Ph.D degree in information science and technology from Osaka University in 2006.
He was a JSPS Research Fellow from 2006-2007.
He was an assistant professor at Osaka University from 2007-2017.
He is now an associate professor of Nara Institute of Science and Technology.
His research interests include program analysis, program comprehension, and software reuse.
He is a member of the IEEE, ACM, IPSJ and JSSST.}

\profile[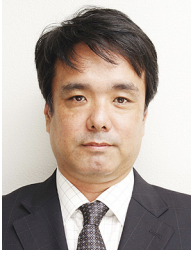]{Kenichi Matsumoto}{received the B.E., M.E., and PhD degrees in Engineering from Osaka University, Japan, in 1985, 1987, 1990, respectively. Dr. Matsumoto is currently a professor in the Graduate School of Information Science at Nara Institute Science and Technology, Japan. His research interests include software measurement and software process. He is a senior member of the IEEE and a member of the IPSJ and SPM.}

\end{document}